\documentstyle[12pt,a4wide,epsfig,amssymb]{article}

\def\SS{\scriptscriptstyle}

\title{
\vspace*{-2.8cm}
\begin{flushright}
{\normalsize DO--TH 97/04\\DOE/ER/01545-709\\OHSTPY-HEP-97-003
\\hep-ph/9701328\\}
\end{flushright}
\vspace{0.8cm}
{\bf Momentum Distribution in the Decay \boldmath $B\rightarrow J/\psi+X$ 
\hspace{-3mm}\unboldmath}
\thanks{ This work was supported in part by the Bundesministerium f\"ur 
Bildung, Wissenschaft, Forschung und Technologie (BMBF), 057DO93P(7), 
Bonn, FRG, and by the US Department of Energy under contract DOE/ER/01545.} 
}
\author{ 
William F. Palmer\\[-1mm]
Department of Physics, The Ohio State University\\[-1mm]
Columbus, OH 43210, USA\\ \\[-1mm] 
Emmanuel A. Paschos, Peter H. Soldan \\[-1mm] 
Institut f\"ur Physik, Universit\"at Dortmund \\[-1mm]
D-44221 Dortmund, Germany 
}
\date{\today}
\begin{document}
\maketitle
\begin{abstract}
We combine the NRQCD formalism for the inclusive color singlet and octet 
production of charmonium states with the parton and the ACCMM model, 
respectively, and calculate the momentum distribution in the decay 
$B\rightarrow J/\psi+X$. Neglecting the kinematics of soft gluon radiation, 
we find that the motion of the $b$ quark in the bound state can account, 
to a large extent, for the observed spectrum. The parton model gives a 
satisfactory presentation of the data, provided that the heavy quark 
momentum distribution is taken to be soft. To be explicit, we obtain 
$\varepsilon_p={\cal O}(0.008-0.012)$ for the parameter of the Peterson 
{\it et al.}~distribution function. The ACCMM model can account for the 
data more accurately. The preferred Fermi momentum 
$p_{\SS F}={\cal O}(0.57$ GeV) is in good agreement with recent studies 
of the heavy quark's kinetic energy. 
\end{abstract}
\newpage
\section{Introduction\label{intro}}
The ARGUS and CLEO collaborations reported results on inclusive $B$ meson 
decays to $J/\psi$, where they identified a sizable component of decays with 
three or more particles in the final state \cite{AaC}. In a recent 
publication \cite{Cleo1} CLEO presented an analysis based on their sample of 
data which is an order of magnitude larger than those of previous studies, 
corresponding to a reduction of errors by a factor of $2.4\,$. As a result the 
group finds the direct branching ratio 
${\cal B}(B\rightarrow J/\psi+X)=(0.80\pm 0.08)\%$. 

In the context of the color singlet (wave function) model this process
is related to the decay of a $b$ quark in the $B$ meson, at 
short distances, into a color singlet $c\bar{c}$ pair plus other quarks and
gluons. The $c$ and $\bar{c}$ quarks generated by this method have almost 
equal momenta and reside in the $^3S_1$ angular momentum state. Several 
authors performed theoretical studies of the branching ratio 
${\cal B}(b\rightarrow J/\psi+X)$, to leading order in $\alpha_s$, using
the color singlet approach \cite{csin}.
The measurements indicate that this approach underestimates the data by 
roughly a factor of three. The inconsistency remains when the next-to-leading 
order perturbative corrections are included \cite{Berg,SoTo}. Thus it is 
interesting to consider generalizations of the color singlet model.

The nonrelativistic QCD (NRQCD) factorization formalism developed by BBL
\cite{BBL} makes possible a systematic treatment of soft gluon effects in
the inclusive heavy quarkonium production. It allows for the creation,
at short distances, of a heavy quark and anti-quark pair in a color octet
configuration which subsequently evolves into a physical bound state through
the emission and absorption of soft gluons.

Color octet contributions to inclusive charmonium production have been 
recently investigated in hadronic reactions at collider and fixed target 
experiments \cite{BrFl,ChL,FM,BRGS}, in $e^+e^-$ annihilations \cite{BrCh},
$\gamma$-nucleon reactions at fixed target and HERA energies 
\cite{CaKr,Am,Ko2}, in $Z^0$ decays at LEP \cite{Z0}, and in lepton-nucleon
reactions \cite{Fl}. A comprehensive review of recent developments in this
field can be found in Ref.~\cite{ccrev}.
In our analysis, we shall refer to previous studies of inclusive 
$B$ meson decays to $J/\psi$ \cite{Ko2,Ko1,FHMN}. These articles
considered the decay of free $b$ quarks, including also the leading color
octet contributions to the branching ratio ${\cal B}(b\rightarrow J/\psi+X)$.
However, the branching ratio is very sensitive to the numerical values chosen 
for the Wilson coefficients and the NRQCD matrix elements. 
In this situation it is desirable to study additional predictions of the
theory. A second comparison deals with the $J/\psi$ momentum distribution. 
Adopting the NRQCD formalism to leading order in the nonrelativistic 
expansion, we shall argue that the observed momentum distribution can be 
attributed, to a large extent, to the $B$ meson bound state corrections. 

So far there is no detailed theoretical fit of the momentum spectrum. 
Palmer and Stech have made a first attempt using a simple wave function 
formalism \cite{Pal-St}. In this paper we investigate two different 
approaches and compare them with the experimental results. As stated above, 
a sizable component in the decay consists of 
nonresonant multi-particle final states. Consequently, for a wide range of
the phase space an inclusive description based on quark-hadron duality is
appropriate. This approach was applied extensively to the inclusive 
semileptonic decays of $B$ mesons.

Over the last few years, considerable progress has been made in the calculation 
of nonperturbative corrections to the lepton energy spectrum of the decay 
$B\rightarrow e\nu+X$ using the operator 
product expansion (OPE) \cite{hqope}. The latter approach 
involves a series in powers of $1/(1-y)m_b$ with $y$ being the normalized 
lepton energy and incorporates the formalism of the Heavy Quark Effective
Theory (HQET) \cite{HQET}. However, an analysis of the decay
$B\rightarrow J/\psi+X$ using the HQET is of limited validity, since
due to the smaller energy release the convergence of the OPE is slower
than for the semileptonic decays.
Therefore we shall resort to a formalism in which the momentum distribution of
the heavy quark in the $B$ meson, the latter representing the dominant
bound state effect, is modeled in terms of parameters  
obtained from experiment. The motion of the $b$ quark will be referred to 
as the smearing of its momentum or the Fermi motion within the $B$ meson.
\\[12pt]\noindent
Several effects may contribute to the momentum spectrum of the $J/\psi$:
\begin{itemize}
\item the motion of the $b$ quark in the bound state, 
\item the shifting of the $c\bar{c}$ momentum due to perturbative 
QCD corrections to the $b\rightarrow c\bar{c}s$ matrix element, carried out 
beyond the leading logarithm approximation (LLA), and
\item the emission or absorption of soft gluons at the hadronization stage.
\end{itemize}
In this work we focus on the dependence of the spectrum on the smearing of 
the $b$ quark momentum in the $B$ meson using the LLA for the short-distance 
matrix element. As far as soft gluon radiation is concerned, we account 
for the color flow using the NRQCD formalism, but neglect the momentum flow 
carried by the soft gluons. This approximation corresponds to keeping the 
leading nonvanishing orders in the $v^2$-expansion for the color singlet 
and octet states, respectively ($v$ being the relative velocity of the 
$c$ and $\bar{c}$). One might note, however, that the $v^2$-expansion fails 
to converge near the boundaries of the phase space. In this region, which is 
sensitive to the mass difference between the partonic $c\bar{c}$ and the 
hadronic $J/\psi$ final state, the momenta of soft gluons become important 
and influence the spectrum \cite{BRW}. 

By analyzing the dependence of the $J/\psi$ momentum
distribution on the initial bound state corrections, we explore one 
of the two relevant nonperturbative effects numerically, keeping in mind 
that the soft gluons, which are radiated within the fragmentation process, 
must be included in the future. To be explicit, we investigate to what extent 
the $b$ quark motion in the $B$ meson can account for the observed spectrum 
in $B\rightarrow J/\psi+X$ decays. 

The paper is organized as follows. In Section \ref{NR} we outline the basic
ideas of the NRQCD approach as they apply to the inclusive charmonium 
production in $b$ decays. We adopt the NRQCD to the decay of a $b\rightarrow
J/\psi+X$, where we emphasize that the $b$ quark is considered to be free 
but the $c\bar{c}$ pairs from both color singlet and octet states convert
into the hadronic $J/\psi$ final state. The latter step is treated in the
leading nonvanishing orders of the $v^2$-expansion for the singlet and octet
states, respectively. In order to account for 
the measured $J/\psi$ momentum spectrum, we shall combine this formalism 
with the $B$ meson bound state corrections, i.e., we consider the
smearing of the $b$ quark momentum in the meson. This we do in two models.
In Section \ref{IPM-sec} we present the parton model which gives a 
satisfactory presentation of the data. Then we calculate in Section 
\ref{AC-sec} the $J/\psi$ momentum spectrum in the framework of the 
ACCMM model \cite{Alta}, which is in better agreement with the data. 
In this context, we take the Fermi motion parameter $p_{\SS F}$ within the 
range obtained from theoretical studies of the $b$ quark's kinetic energy. 
Our analysis confirms that, by including the leading color octet contributions
to the charmonium production, one can account for the measured branching
ratio. In addition, we find that the $b$ quark motion gives a good
representation of the spectrum. The summary and the discussion of our
results can be found in Section \ref{summary}.

\section{\boldmath $b\rightarrow J/\psi+X$ \unboldmath in the NRQCD
{}Formalism \label{NR} }

The nonrelativistic-QCD (NRQCD) formalism, developed by BBL \cite{BBL},
represents a comprehensive theoretical framework for the study of inclusive
charmonium production.
Denoting by $v$ the relative velocity of the two heavy constituents inside 
the $(Q\bar{Q})$ bound state, the NRQCD approach introduces the complete
structure of the quarkonium Fock space to a given order in $v^2$ and thus 
allows for a consistent factorization of long- and short-distance effects.
Consequently, the charmonium production rate is represented by a sum of
products, each of which consists  of a short-distance coefficient, 
associated with the creation of a heavy quark and anti-quark pair in a 
specific angular and color configuration, and a nonperturbative NRQCD matrix 
element $\langle 0|O^H(^{2S+1}L_J)_a|0\rangle$ which parameterizes the 
subsequent evolution of the intermediate $c\bar{c}(^{2S+1}L_J)_a$ state
into a physical charmonium bound state $H$ (plus light hadrons). 
The latter step is generated through the emission and absorption of soft 
gluons.

The separation of the distance scales (see Fig.~1), and its association to 
the small value of the constituents' relative velocity ($v_c^2\sim 0.23-0.30$ 
for charmonium), makes possible the calculation of the heavy quarkonium 
production rate in terms of a double series in powers of $v^2$ and 
$\alpha_s$ (numerically $\alpha_s(m_c^2)\sim v_c^2$). This way the 
production through an intermediate color octet state, although down by 
powers of $v^2$ in the Fock state hierarchy, can be numerically relevant 
in view of possible short-distance enhancements.
The authors of Refs.~\cite{Ko2,Ko1,FHMN} pointed out that, for inclusive $B$
decays to S-wave charmonia, the color octet production mechanism is indeed
important, even though it is of order $v^4$ with respect to the basic 
color singlet one, because the Wilson coefficient of the color octet term is
strongly enhanced in comparison to that of the color singlet transition.
Since we shall use formulas for the decay rate $\Gamma(b\rightarrow J/\psi+X)$
in our analysis of the ACCMM model, we briefly outline the basic ideas of 
the NRQCD approach.\\[12pt]
At the $b$ mass scale, the Hamiltonian which induces the effective 
$b\rightarrow c\bar{c}q_f$ $(f=s,\,d)$ transitions reads
\begin{equation}
{\cal H}_{ef\hspace{-0.5mm}f}=\frac{G_F}{\sqrt{2}}V^{ }_{cb}V_{cf}^*
\left[ \frac{1}{3}(2C_+-C_-)\bar{c}\gamma_\mu Lc\,\bar{q}_f
\gamma^\mu L b
+(C_++C_-)\bar{c}\gamma_\mu L T^a
c\,\bar{q}_f\gamma^\mu L T^a b \right]\,,   
\label{Heff}
\end{equation}
$L=1-\gamma_5$,
where operators arising from penguin and box diagrams have been neglected. 
The renormalization (Wilson) coefficients $C_\pm(\mu)$ have been 
computed up to the next-to-leading order corrections in Ref.~\cite{Buras}.

In the leading logarithm approximation (LLA), the decay amplitude 
refers to the tree level matrix element of the effective Hamiltonian. 
Consequently, the renormalization scale dependence contained in the Wilson 
coefficient functions, $C_\pm(\mu)$, cannot be cancelled by this matrix 
element. While this work was being completed, a study of higher order 
perturbative corrections to the matrix element of $b\rightarrow J/\psi+X$ 
decays appeared \cite{SoTo}, in which a double series expansion in 
$\alpha_s$ and the small ratio of the (LL) color singlet to octet Wilson 
coefficients was performed. Using the color singlet approximation for 
charmonium production, the authors obtained a result which carries only weak 
dependence on the renormalization scale. The predicted branching ratio, 
 ${\cal B}(B\rightarrow J/\psi+X)=0.9{+1.1\atop -0.3}\times 10^{-3}$
lies well below the experimental value of $(0.80\pm 0.08)\%$ reported
by the CLEO collaboration. This confirms the supposition that a 
nonperturbative effect in the $B\rightarrow J/\psi+X$ transition 
is required, which goes beyond the color singlet approximation,
namely the color octet (NRQCD) mechanism for charmonium production. 
In future studies the leading order result for the color singlet 
contribution can be replaced by the expression given in Ref.~\cite{SoTo}. In 
this article we keep the leading logarithm approximation, assuming that the 
modification of the $J/\psi$ momentum distribution will be small.\\[12pt] 
\noindent
In order to obtain the nonrelativistic interaction from the effective
Hamiltonian of Eq.~(\ref{Heff}), one expresses the 
Dirac bilinears of the $c$ and $\bar{c}$ in terms of (two-component) 
Pauli spinors $\xi$ and $\eta$, the constituents' relative three-momentum 
$\bf{q}$, and the Pauli matrices $\sigma_i$. This allows one to consider 
separately states with specific angular momentum and color (for details see
Ref.~\cite{FM}). To linear order in $v=|{\bf q}|/m_c$, the short-distance 
amplitude $\cal A$ (using a notation in which the spinors do
not carry color indices) can be rewritten as \cite{FHMN} 
\begin{eqnarray}
{\cal A}(\sigma,\tau;c,d,e,f;\mbox{\bf q})&=&\frac{G_F}{\sqrt
2}V_{cb}^{}V_{cs}^*\left[\frac{1}{3}(2C_+-C_-)\delta^{cd}\delta^{ef}+
(C_++C_-)T^g_{cd}T^g_{ef}\right]\label{ME}\\&&
\times\bar{s}\gamma^\mu Lb \hspace*{1.5mm}\xi_\sigma^\dagger
\left[2m_c\Lambda_i^\mu\sigma^i-P^\mu+2i\Lambda_k^\mu\varepsilon^{mik}
q_m\sigma_i\right]\eta_\tau\nonumber\,,
\end{eqnarray} 
$c,d,e,f$ being color indices, $\sigma$ and $\tau$ the individual spins of
the charm quarks. $P$ is the total four-momentum of the $c$ and $\bar{c}$
in the laboratory frame, $\Lambda_i^\mu$ is the Lorentz boost matrix that 
takes a three-vector from the $c\bar{c}$ restframe to the laboratory frame.

The decay rate to a $c\bar{c}$ pair, calculated in full perturbative QCD, 
is obtained from 
\begin{equation}
\Gamma(b\rightarrow c\bar{c}+q_f)=\frac{1}{2E_b}\int\frac{d^3{\bf q}}
{(2\pi)^3}\frac{d^3{\bf P}}{(2\pi)^3 2P_0}\frac{d^3{\bf p}_f}{(2\pi)^3 2E_f}
(2\pi)^4\delta^{(4)}(p_b-p_f-P)|\bar{\cal A}|^2 m_c^{-1}+\ldots\,\,,
\label{FPQCD}
\end{equation}
where the dots denote higher order terms in the $v^2$-expansion.
The integrations represent the three-body phase space of the final state.
Using perturbative NRQCD, the corresponding decay rate to a $c\bar{c}$
pair of a specific angular and color quantum number state 
$[n]_a\equiv (^{2S+1}L_J)_a$ reads
\begin{equation}
\Gamma(b\rightarrow c{\bar c}[n]_a+q_f)=\int\frac{d^3 {\bf q}}{(2\pi)^3}
S_a[n]\frac{\langle 0|O_a^{c\bar{c}}[n]|0\rangle}{m_c^2}+\ldots\,\,.
\label{PNRQCD}
\end{equation}   
Replacing the heavy quark field operators, which appear in the $c\bar{c}$ 
matrix elements, in terms of the Pauli spinors $\xi$ and $\eta$ permits a 
matching of the two perturbative results in Eqs.~(\ref{FPQCD}) and 
(\ref{PNRQCD}).  This determines the short-distance coefficients $S_a[n]$, 
which equally apply to the inclusive decays to charmonium {\it bound} states 
(for more details see Refs.~\cite{ChL,FM}).

Consequently, the rate of the transition $b\rightarrow J/\psi+X$
in the NRQCD factorization formalism reads
\begin{eqnarray}
\Gamma(b\rightarrow J/\psi+X)&=&
\sum_{[n]_a}S_a[n]\frac{\langle 0|O_a^{J/\psi}[n]|0\rangle}{m_c^2}\\
&=&\frac{1}{2 E_b}\int\frac{d^3{\bf P}}{(2\pi)^3 2P_0}\frac{d^3{\bf p}_f}
{(2\pi)^3 2E_f}(2\pi)^4\delta^{(4)}(p_b-p_f-P)m_c^{-1} \nonumber \\
&&\times \sum_{[n]_a}
\frac{|{\bar{\cal A}}|^2(b\rightarrow c\bar{c}[n]_a+q_f)}
{\langle 0|O_a^{c\bar{c}}[n]|0\rangle}\langle 0|O_a^{J/\psi}[n]|0\rangle\,.
\label{NRQCD}
\end{eqnarray}
Note that the integration over the relative momentum in Eq.~(\ref{PNRQCD})
properly relates the short-distance $1\rightarrow 3$ particle transition to
the long-distance effective $1\rightarrow 2$ particle one.

Evaluating the color octet intermediate states in Eq.~(\ref{NRQCD}) 
up to order $v^4$ relative to the color singlet `baseline', one obtains in
the restframe of the $b$ meson (with $m_f\simeq 0$ and
$|V_{cs}|^2+|V_{cd}|^2\simeq 1$) \cite{Ko2,Ko1,FHMN},
\begin{equation}
\Gamma (b\rightarrow J/\psi+X) =
\frac{G_F^2}{144\pi}|V_{cb}|^2 m_c m_b^3\left(1-\frac{4m_c^2}{m_b^2}
\right)^2\left[a\left(1+\frac{8m_c^2}{m_b^2}\right)+b\right]\,,  
\label{br0}
\end{equation}
where we defined the NRQCD coefficients
\begin{eqnarray}
a&=&(2C_+-C_-)^2
\frac{\langle 0|O_1^{J/\psi}( ^3S_1)|0\rangle}{3m_c^2}\label{a}\\
&&+(C_++C_-)^2\left[
\frac{\langle 0|O_8^{J/\psi}( ^3S_1)|0\rangle}{2m_c^2}+
\frac{\langle 0|O_8^{J/\psi}( ^3P_1)|0\rangle}{m_c^4}\right]\,,\nonumber \\
b&=&(C_++C_-)^2\frac{3\langle 0|O_8^{J/\psi}( ^1S_0)|0\rangle}{2m_c^2}
\label{b}\,.
\end{eqnarray}
We shall apply Eqs.~(\ref{br0})-(\ref{b}) in Section 4 in the context of the 
ACCMM model in order to determine the $J/\psi$ momentum distribution from 
$B$ meson decays.\\[12pt]
\noindent
In the case of the parton model, which we shall introduce in Section 3, we 
have to relate the short- and long-distance matrix elements squared rather 
than the decay rates.
To this end, let us define the matrix element $\cal M$ of the
effective interaction $b\rightarrow J/\psi+X$, in which the $b\rightarrow
q_f$ transition is assumed to be independent of the inclusive charmonium
production process, through 
\begin{equation}
\Gamma(b\rightarrow J/\psi+X)=
\frac{1}{2 E_b}\int\frac{d^3{\bf k}_\psi}{(2\pi)^3 2E_\psi}
\frac{d^3{\bf p}_f}{(2\pi)^3 2E_f}(2\pi)^4\delta^{(4)}(p_b-p_f-k_\psi)
|\bar{\cal M}|^2\,.
\label{breff}
\end{equation}
If we identify the total four-momentum $P$ of the $c$ and $\bar{c}$ with the 
momentum of the $J/\psi$ in the final state, $k_\psi\equiv P$, then 
comparing Eqs.~(\ref{NRQCD}) and (\ref{breff}) we obtain
\begin{equation}
|\bar{\cal M}|^2(b\rightarrow J/\psi+X)=m_c^{-1}\sum_{[n]_a}
\frac{|\bar{\cal A}|^2(b\rightarrow c\bar{c}[n]_a+q_f)}
{\langle 0|O_a^{c\bar{c}}[n]|0\rangle}
\langle 0|O_a^{J/\psi}[n]|0\rangle\,.
\end{equation}
An explicit calculation yields
\begin{equation}
\bar{|{\cal M}|^2}=G_F^2 |V_{cb}|^2 T_{\mu\nu}
\langle(p_f\hspace{-3.7mm}/\hspace{+1.1mm}+m_f)\gamma^\mu L
(p_b\hspace{-3.2mm}/\hspace{+1.0mm}+m_b)\gamma^\nu L \rangle\,,
\end{equation} 
where the $J/\psi$ tensor structure $T_{\mu\nu}$ reads 
\begin{equation}
T_{\mu\nu}=|V_{cf}|^2m_c^3\frac{1}{9}\left[-ag_{\mu\nu}
+(a+b)\frac{1}{M_\psi^2}(k_\psi)_\mu (k_\psi)_\nu\right]\,,
\label{psiten}
\end{equation}
with $a$ and $b$ defined in Eqs.~(\ref{a}) and (\ref{b}).
The momentum identification, described above, is in accordance with the BBL 
formalism for quarkonium production. To leading order in the 
$v^2$-expansion, the soft gluons, radiated in the nonperturbative transition 
$c\bar{c}\rightarrow J/\psi+X$ (and materializing into light hadrons), are 
assigned no energy or momentum; i.e., the energy dependence of the 
long-distance interaction is neglected, and thus the NRQCD matrix elements 
appear as universal constants.

In Section 3 we shall apply Eq.~(\ref{psiten}) in the context of the parton
model. For comparison with data we identify $2m_c$ with the mass $M_\psi$
of the $J/\psi$, because to leading order in the $v^2$-expansion $P^2=4m_c^2$
(however, one might keep in mind that the expansion fails to converge
near the endpoint of the spectrum).  
\section{\boldmath $B\rightarrow J/\psi+X$ \unboldmath
in the Parton Model (PM) \label{IPM-sec}}
\subsection{Calculation of the Differential Branching Ratio}
As we mentioned above, in the context of the NRQCD approach the
nonperturbative charmonium production process $c\bar{c}\rightarrow
J/\psi+X$ is assumed to be independent of the $b\rightarrow q_f$ transition.
Thus, within the leading logarithm approximation for the short-distance
matrix element, the $B\rightarrow J/\psi +X$ amplitude can be evaluated using 
a (generalized) factorization approach; i.e., the amplitude is related to a 
product of matrix elements of current operators \cite{HS},  
\begin{eqnarray}
{\cal M}(B\rightarrow J/\psi+X_f)
&=&\frac{G_F}{\sqrt 2}V_{cb}^{}V_{cf}^*
\Bigg[\frac{1}{3}(2C_+-C_-)
\langle X_f^1|\bar{q}_f\gamma_\mu L b|B\rangle
\langle\psi+X^1|\bar{c}\gamma^\mu L c|0\rangle\nonumber \\
&&+(C_++C_-)
\langle X_f^8|\bar{q}_f\gamma_\mu L T^ab|B\rangle
\langle\psi+X^8|\bar{c}\gamma^\mu L T^ac|0\rangle\Bigg]\,,
\label{mel}
\end{eqnarray}
with $X_f^1+X^1=X_f$, $X_f^8+X^8=X'_f$.
Note that, contrary to the color singlet wave function model, 
Eq.~(\ref{mel}) includes the color octet transitions.

The standard NRQCD approach outlined in Section 2 considers charmonium 
production from a {\it free} $b$ quark decay. As it becomes obvious 
from Eq.~(\ref{NRQCD}), the corresponding transition $b\rightarrow J/\psi+X$ 
refers to a $1\rightarrow 2$ particle phase space, because we neglect the soft 
gluon momenta. In this approximation, the $J/\psi$ momentum 
distribution from $B$ meson decays arises from the initial bound 
state corrections (and the momentum smearing from the Lorentz boost to the 
laboratory frame). We incorporate the corrections, attributed to the 
interaction between heavy and light quark in the initial meson, in the 
parton model along the lines developed for semileptonic $B$ decays \cite{Jin1}.

If we use Eq.~(\ref{mel}), which is valid at the level of tree calculations, 
in the restframe of the $B$ we can write the decay rate as
\begin{equation}
d\Gamma{(B\rightarrow J/\psi+X_f)}=\frac{G_F^2 |V_{cb}|^2}{(2\pi)^2M_B}
T_{\mu\nu}W^{\mu\nu}\frac{d^3{\bf k}_\psi}{2E_\psi}\,,
\end{equation} 
with $T_{\mu\nu}$ defined in Eq.~(13). 
Introducing the light-cone dominance, as in Ref.~\cite{Jin1}, allows us to 
relate the transition $B\rightarrow X_f$ to the distribution function $f(x)$ 
of the heavy quark momentum,
\begin{equation}
W_{\mu\nu}=4(S_{\mu\rho\nu\lambda}-i\varepsilon_{\mu\rho\nu\lambda})
\int_0^1 dx\,f(x)P_B^\lambda(xP_B-k_\psi)^\rho \varepsilon[(xP_B-k_\psi)_0]
\delta[(xP_B-k_\psi)^2-m_f^2]\, ,
\label{tensor2}
\end{equation} 
with 
\begin{equation}
S_{\mu\rho\nu\lambda}=g_{\mu\rho}g_{\nu\lambda}-g_{\mu\nu}g_{\rho\lambda}
+g_{\mu\lambda}g_{\nu\rho} 
\hspace{5mm} \mbox{and} \hspace{5mm}
\varepsilon (x)=\left\{ \begin{array}{ll}
   +1\, , \hspace{4mm} & x > 0 \\ -1 \, , & x < 0 \, . \end{array}  \right.  
\end{equation} 
The dependence of the distribution function on the single scaling variable 
$x$ is a consequence of the light-cone dominance, since in this framework 
the structure function $f(x)$ is obtained as the Fourier transform of the
reduced bilocal matrix element between the hadronic states \cite{Jin3},
\begin{equation}
f(x)=\int d(yP_B)e^{ix(yP_B)}\frac{1}{4\pi M_B^2}
\langle B|\bar{b}(0)P_B^\mu \gamma_\mu^L b(y)|B\rangle|_{y^2=0} \, .
\label{disfunc}
\end{equation} 
The Lorentz invariant width of the decay reads
\begin{eqnarray}
E_B\cdot d\Gamma&=&\sum_{f=s,\,d}\frac{C_f}{\pi}\int dx\, f(x)
\varepsilon[(xP_B-k_\psi)_0]\delta^{(1)}\left[(xP_B-k_\psi)^2-m_f^2\right]
\label{diffbr}\\
&&\times\left[ (a-b)P_B(xP_B-k_\psi)+\frac{2}{M_\psi^2}(P_Bk_\psi)
(xP_B k_\psi-M_\psi^2)(a+b)\right]\frac{d^3{\bf k}_\psi}{2 E_\psi} \,,  
\nonumber
\end{eqnarray}
where
\begin{equation}
C_f=\frac{G_F^2}{72\pi}|V_{cb}|^2|V_{cf}|^2M_\psi^3\,.
\end{equation}
Evaluating Eq.~(\ref{diffbr}) in the restframe of the $B$ meson, we arrive at
the formula for the $J/\psi$ momentum spectrum,
\begin{eqnarray}
\lefteqn{\frac{1}{\Gamma_B}\frac{d\Gamma}{d|\mbox{\bf k}_{\psi}|}
(B\rightarrow J/\psi+X)\,=\,\tau_B\frac{G_F^2}{72\pi M_B}|V_{cb}|^2M_\psi^3
\frac{|\mbox{\bf k}_{\psi}|^2}{E_\psi}}
\label{sim}
\\&&\times
\left\{\left[f(x_+)+f(x_-)\right]\left[(a-b)+(a+b)\frac{2E_\psi^2}{M_{\psi}^2}
\right]+\left[f(x_+)-f(x_-)\right](a+b)\frac{2E_\psi 
|\mbox{\bf k}_\psi|}{M_{\psi}^2}
\right\}\,,
\nonumber
\end{eqnarray}
within the kinematical range 
\begin{equation}
M_{\psi}\le E_{\psi}\le \frac{M_B^2+M_{\psi}^2-m_f^2}{2M_B}\,.
\label{kin}
\end{equation}
Here we defined
\begin{equation}
x_{\pm}  =  \frac{1}{M_B} 
\left( E_{\psi} \pm \sqrt{|k_{\psi}|^2+m^2_f}\right)
\end{equation}
and adopted $m_f\simeq 0$ and $|V_{cs}|^2+|V_{cd}|^2\simeq 1$.
The kinematical range for $x_-$ belongs to a final state quark with negative 
energy. Therefore the corresponding terms can be associated, formally, with
quark pair-creation in the $B$ meson whereas the dominant terms proportional 
to $f(x_+)$ reflect the direct decay.

To compare our result with data from CLEO a Lorentz boost has  
in addition to be performed from the restframe of the $B$ meson
to a moving $B$ produced at the $\Upsilon (4S)$ resonance 
($|{\bf p}_B|=0.34$ GeV). The explicit form of the boost integral is 
presented in Section \ref{ac-cal} in the context of the ACCMM model 
(see Eq.~(\ref{brac})).\\[12pt]
We note that the dependence of the decay spectrum on the structure function 
appears in a factorized form. Thus the shape of the spectrum in  
Eq.~(\ref{sim}) is governed by the functional form of the heavy quark
momentum distribution. This is different for the 
semileptonic decay, in which an integral over the structure function is
involved. In the following section we compare the predictions of  
Eq.~(\ref{sim}) with the existing experimental data.
%
\subsection{Analysis and Numerical Evaluation \label{pman}}
In order to compute the theoretical momentum distribution we must fix the
parameters appearing in Eq.~(\ref{sim}). In the numerical analysis we
use $G_F=1.1664\times 10^{-5}$ GeV$^{-2}$, $M_B=5.279$ GeV, 
$M_{\psi}=3.097$ GeV and $C_+(m_b)=0.868$, $C_-(m_b)=1.329$. The
(leading log) values of the Wilson coefficients correspond to the
central value of the pole mass, $m_b=4.7$ GeV \cite{mb}.
Moreover we adopt $\tau_B=1.61$ ps for the $B$ meson lifetime
\cite{tB} and $|V_{cb}|=0.040$. The latter value stems from a CLEO II
analysis \cite{vcbcl} of the inclusive semileptonic $B$ decays 
\footnote{The analysis was based on a modified version of the ISGW model
\cite{ISGW} and on ACCMM \cite{Alta}, respectively, using $p_{\SS F}=0.3$
GeV for the Fermi momentum parameter. The corresponding uncertainty, among
other things due to the ambiguity in the choice of $p_{\SS F}$, we postpone
to the discussion of the overall uncertainty associated with the
normalization of the $B\rightarrow J/\psi +X$ decay spectrum.}
and is in accordance with the world average from inclusive and 
exclusive measurements, $|V_{cb}|=0.0381\pm 0.0021$ \cite{vcbav}.

As we mentioned above, the numerical values of the nonperturbative NRQCD
matrix elements $\langle 0|O_a^{J/\psi}(^{2S+1}L_J)|0\rangle\equiv \langle
O_a^{J/\psi}(^{2S+1}L_J)\rangle$ must be determined phenomenologically.
For the rest of this article, we use\begin{equation}
\begin{array}{lcl}
\langle O_1^{J/\psi}(^3S_1)\rangle &=&\hspace*{3.3mm}1.1\,\,\mbox{GeV}^3
\,,\nonumber\\[0.1cm]
\langle O_8^{J/\psi}(^3S_1)\rangle &=&\hspace*{3.3mm}0.0066\,\,\mbox{GeV}^3
\,,\nonumber\\[0.1cm]
\langle O_8^{J/\psi}(^1S_0)\rangle &=&\hspace*{3.3mm}0.04\,\,\mbox{GeV}^3
\,,\nonumber\\[0.1cm]
\langle O_8^{J/\psi}(^3P_0)\rangle /m_c^2&=&-0.003\,\,\mbox{GeV}^3\,.
\label{meval}
\end{array}
\end{equation}
In addition we have to impose the (approximate) heavy quark spin symmetry
to relate the matrix element $\langle O_8^{J/\psi}(^3P_0)\rangle$ to that 
arising from the $(^3P_1)$ intermediate state relevant for $B$ decays,
\begin{equation}
\langle O_8^{J/\psi}(^3P_J)\rangle \simeq (2J+1)
\langle O_8^{J/\psi}(^3P_0)\rangle \,.
\end{equation}
The color singlet matrix element $\langle O_1^{J/\psi}(^3S_1)\rangle$
has been determined from the leptonic width of the $J/\psi$ \cite{BSK}, the
corresponding color octet matrix element $\langle O_8^{J/\psi}(^3S_1)\rangle$
stems from the analysis of $J/\psi$ hadroproduction at CDF \cite{ChL}.
The remaining parameters $\langle O_8^{J/\psi}(^1S_0)\rangle$ and
$\langle O_8^{J/\psi}(^3P_0)\rangle$ have been extracted from a fit to
data on $J/\psi$ leptoproduction \cite{Fl} (the latter involving only a
small dependence on the value chosen for $m_c$).

Note that the renormalized matrix element
$\langle O_8^{J/\psi}(^3P_0)\rangle$ is not restricted to positive values, a
characteristic due to the subtraction of large power divergences from the
(positive definite) bare matrix element (for details see Ref.~\cite{Fl}).

The numerical values of $\langle O_8^{J/\psi}(^1S_0)\rangle$ and
$\langle O_8^{J/\psi}(^3P_0)\rangle$ quoted in Eq.~(\ref{meval}) are
consistent with results from photoproduction \cite{Am}, pion-nucleon
reactions \cite{BRGS} and, roughly, with calculations of $J/\psi$
production in hadronic collisions \cite{ChL}.
Yet one has to note that the various investigations of charmonium
production processes within the NRQCD approach involve leading order 
theoretical expressions. The corresponding uncertainties in the determination 
of the matrix elements set up a sizable range in the parameter space
(for a detailed discussion see Ref.~\cite{FHMN}).
However, as we discuss next, an important feature of our analysis is 
that the momentum dependence is largely free of these uncertainties.

In both models, the parton and the ACCMM model, the shapes of the 
spectra originating from various $c\bar{c}(^{2S+1}L_J)$ intermediate states 
are essentially identical. This statement is obvious in case of the 
$(^3L_1)$ vector states, which involve identical Lorentz structures, and thus
the relevant NRQCD matrix elements appear in a single constant (see
Eq.~(\ref{a})); in case of the $(^1S_0)$ scalar state the equivalence
follows from the explicit calculation.   
Consequently, the shape of the entire $J/\psi$ momentum distribution in
$B$ meson decays is practically independent of the exact values of the 
nonperturbative matrix elements. The same statement holds for the numerical
values of the parameters occurring in the short-distance coefficients (in 
particular errors have to be attributed to the Wilson coefficients, due to
scale uncertainties, and to the value chosen for the CKM element $V_{cb}$).

The set of parameters mentioned above solely determines the normalization
of the $J/\psi$ momentum spectrum. Within our approach the shape is attributed
to the momentum distribution of the $b$ quark in the initial $B$ meson. 
Further investigation of soft gluon effects is clearly needed. The first 
step we take here is to ask: can a successful reproduction of the data at 
hand be achieved without them.\\[12pt]
In the parton model, in the absence of direct measurements of the 
distribution function we use a one-parameter Ansatz and fix the distribution 
parameter by comparing our results with data. Referring to theoretical
studies which pointed out that the distribution and fragmentation function
of heavy quarks peak at large values of $x$ \cite{BjBr}, we assume, as a 
working hypothesis, that the functional form of both is similar. The
latter is known from experiment and we shall use the Peterson functional
form \cite{PCB}
\begin{equation}
f(x)=N_\varepsilon
\frac{x(1-x)^2}{[(1-x)^2+\varepsilon_p x]^2}\, ,
\label{peterson}
\end{equation}
with $\varepsilon_p$ being the free parameter and  $N_\varepsilon$ the
corresponding normalization constant. This function has already been
applied in the semileptonic decays of the $B$ meson \cite{Jin1}.

In Fig.~2a we show the distribution function $f(x)$ for various values of
$\varepsilon_p$. The kinematical range for the two arguments $x_\pm$
appearing in Eq.~(\ref{sim}) reads
\begin{equation}
\frac{M_{\psi}}{M_B}\le x_+\le 1\, ,
\qquad \frac{M^2_{\psi}}{M^2_B} \le x_- \le \frac{M_{\psi}}{M_B}\, .
\end{equation}
Note that the variable $x_-$ only occurs with values at which
$f(x_-)$ is small. Therefore the corresponding contribution
to the differential branching ratio is small.

We use Eq.~(\ref{peterson}) to fit the measured momentum spectrum of the
$J/\psi$ which was presented by the CLEO group \cite{Cleo1}. As pointed out
above, the shape of the theoretical spectrum is determined by the
distribution function, i.e., by the parameter $\varepsilon_p$.

A general feature of the analysis is the difficulty for reproducing the data
over the entire range of phase space. Confronted with this problem, we lay
greater emphasis on the appropriate description of the {\it low} momentum
range ($|\mbox{\bf k}_\psi| \lesssim 1.4$ GeV).
Within this region the $J/\psi$ spectrum obtains a sizable contribution
from decay channels containing three or more particles in the final state, 
whereas the high momentum range is mainly determined by the exclusive 
two-body decays $B\rightarrow J/\psi\,K^{(*)}$. Therefore incoherence, as a 
necessary ingredient of the PM, is fulfilled in the former region.
In addition, the $v^2$-expansion fails to converge in the endpoint domain
of the spectrum. 

As a result we find that if we apply small values for the 
parameter $\varepsilon_p\lesssim 0.004- 0.006$
we cannot account for any part of the low momentum region which is 
underestimated, whereas the high momentum range is overestimated. 
The situation improves if we apply larger values for 
$\varepsilon_p={\cal O}(0.008-0.012)$, corresponding to a soft $b$ quark 
distribution (see Fig.~3). For $\varepsilon_p=0.008$ we obtain a first
satisfactory fit for the present data. Yet there is still a moderate 
systematic underestimate. A further increase of the distribution parameter  
(with the remaining parameters kept unchanged) does not imply significant  
modifications within the low momentum range. For comparison, one might note 
that in the studies of semileptonic $B$ decays $\varepsilon_p$ was taken 
between $0.003$ and $0.009$ \cite{Jin1}. 

Considering the uncertainties associated with the determination of the NRQCD
matrix elements and, consequently, with the normalization of the spectrum,
in Fig.~4 we show the $J/\psi$ momentum distribution for $\varepsilon_p=0.012$ 
and a modified value 
$\langle O_8^{J/\psi}(^3P_0)\rangle/m_c^2=-0.001$ GeV$^3$. 
The fit is somewhat better. However, the color octet intermediate states, 
with reasonable values of the matrix elements, improve the fit of the 
calculated curves to the observed spectrum. 
{}For comparison, and in order to point out the importance of the color 
octet charmonium production mechanism, we also show in Fig.~4 the spectrum 
arising from the color singlet (wave function) model 
($\langle O_8^{J/\psi}(^{2S+1}L_J)\rangle\equiv 0 $). 
The latter underestimates the data by roughly a factor of three.  

As a general result we conclude that a soft $b$ quark momentum 
distribution can, to a good extent, account for the observed spectrum
in $B\rightarrow J/\psi+X$ decays. We shall see in Section \ref{ac-an} than 
an analogous statement holds for the Fermi momentum distribution in the 
ACCMM model. 

{}Finally, the errors in the data, although substantially improved, are still
significant and a better test will be possible, when the error bars will be
further reduced.  With better data on should also include higher order
nonperturbative corrections to the quarkonium production, and study the 
influence of soft gluon radiation within the fragmentation process.

\section{\boldmath $B\rightarrow J/\psi+X$ \unboldmath
in the ACCMM Model \label{AC-sec}}
\subsection{Calculation of the Differential Branching Ratio \label{ac-cal}}

A second approach which allows us to analyze the momentum distribution of
$J/\psi$ in the inclusive decay of the $B$ meson is given by
the ACCMM model \cite{Alta}. In this model the bound state corrections to 
the free $b$ quark decay are incorporated by attributing to the spectator
quark a Fermi motion within the meson. The momentum spectrum of the 
$J/\psi$ is then obtained by folding the Fermi motion with the spectrum from
the $b$ quark decay. 
In Ref.~\cite{Barger} the shape of the $J/\psi$ momentum distribution 
resulting from Fermi momentum smearing has been given in the context of the
color singlet wave function model (and without consideration of the light
spectator quark mass). We shall extend the analysis by including the leading 
color octet contribution to the charmonium production in the framework of 
the NRQCD factorization formalism and by comparing our results with
experimental data.

The spectator quark is handled as an on-shell particle with definite mass
$m_{sp}$ and momentum $|{\bf p}|=p$.
Consequently, the $b$ quark is considered to be off-shell with a
virtual mass $W$ given in the restframe of the $B$ meson
by energy-momentum conservation as
\begin{equation}
W^2(p)=M_B^2+m_{sp}^2-2M_B\sqrt{m_{sp}^2+p^2}\,. \label{vmass}
\end{equation}
{\it Altarelli et al.} introduced in the model a Gaussian probability
distribution $\phi(p)$ for the spectator (and thus for the heavy 
quark) momentum, 
\begin{equation}
\phi(p)=\frac{4}{\sqrt{\pi}p_{\SS F}^3}\exp\left(-p^2/p_{\SS F}^2\right) \, ,
\label{ac-dist}     
\end{equation}
normalized according to 
\begin{equation}
\int_0^\infty dp\, p^2 \phi(p) = 1\, .
\end{equation}
The Gaussian width $p_{\SS F}$ is treated as a free parameter which has to
be determined by experiment.

One main difference between the parton model and ACCMM is that in
the latter one must consider a $b$ quark in flight.   
We therefore start from the momentum spectrum of the $J/\psi$ 
resulting from the decay 
$b\rightarrow J/\psi+X_f$ $(f=s,\,d)$ 
of a $b$ quark of mass $W$ and momentum $p$ which is given by     
\begin{equation}
\frac{d\Gamma_b}{d|{\bf k}_{\psi}|}\left(|{\bf k}_{\psi}|,\,p\right)=   
\gamma_b^{-1}
\frac{\Gamma_0}{k_+^{(b)}(p)-|k_-^{(b)}(p)|}\left[\theta 
\left(|{\bf k}_{\psi}|-|k_-^{(b)}(p)|\right)-\theta 
\left(|{\bf k}_{\psi}|-k_+^{(b)}(p)\right)\right] \, .
\label{gbp}
\end{equation}
Here we have defined 
\begin{equation}
\theta (x)=\left\{ \begin{array}{ll}
   1, \hspace{4mm} & x > 0 \\ 0, & x < 0 \, . \end{array}  \right. 
\end{equation}
$\Gamma_0$ is the width of the analogous decay in the restframe of the
heavy quark as obtained from the NRQCD factorization formalism. According
to Eqs. (\ref{br0})-(\ref{b}) it reads (again adopting $m_f\simeq 0$
and  $|V_{cs}|^2+|V_{cd}|^2\simeq 1$) 
\begin{equation}
\Gamma_0=\frac{G_F^2}{144\pi}|V_{cb}|^2 m_c W^3\left(1-\frac{4m_c^2}{W^2}
\right)^2\left[a\left(1+\frac{8m_c^2}{W^2}\right)+b\right]\,,  
\end{equation}
where $a$ and $b$ contain the nonperturbative effects in the charmonium
production process.

In Eq.~(\ref{gbp}) $k^{(b)}_\pm$ give the limits of the momentum range
which results from the Lorentz boost from the restframe of the $b$ 
quark to a frame where the $b$ has a nonvanishing momentum $p$,
\begin{equation}
k^{(b)}_{\pm}(p)=\frac{1}{W}\left(E_b k_0\pm p E_0 \right)\, ,
\end{equation}
with
\begin{equation}
k_0=\frac{1}{2W}\left(W^2-M_{\psi}^2\right) \;, \hspace{5mm} 
E_0= \sqrt{k_0^2+M_{\psi}^2} \,, 
\end{equation}
and $\gamma_b^{-1}$ being the corresponding Lorentz factor, 
\begin{equation}
\gamma_b^{-1}=\frac{W}{E_b}\;, \hspace{1cm} E_b=\sqrt{W^2+p^2}\,.
\end{equation}
To calculate the momentum spectrum of the $J/\psi$ from the 
inclusive decay of the $B$ meson one has to fold the heavy quark momentum 
probability distribution with the spectrum of Eq.~(\ref{gbp}) resulting 
from the $b$ quark subprocess. Performing this, we finally arrive at the 
expression for the differential branching ratio for a $B$ meson in flight,  
\begin{eqnarray}
\lefteqn{\frac{1}{\Gamma_B}
\frac{d\Gamma}{d|{\bf k}_{\psi}|}\left(B\rightarrow J/\psi+X\right)= }
\label{brac}\\
&&\hspace*{1cm}
\tau_B \int\limits_{|k_-(|{\bf k}_{\psi}|)|}^{\hat{k}_+(|{\bf k}_{\psi}|)}\;
\frac{d|{\bf k}_{\psi}'|}{k_+(|{\bf k}_{\psi}'|)-|k_-(|{\bf k}_{\psi}'|)|}
\int\limits_0^{p_{max}}dp\;p^2\phi(p)
\frac{d\Gamma_b}{d|{\bf k}_{\psi}|}\left(|{\bf k}_{\psi}'|,\,p\right) \, .
\nonumber
\end{eqnarray}
Here $p_{max}$ is the maximum kinematically allowed value of the quark
momentum $p$, i.e., that which makes $W$ in Eq.~(\ref{vmass}) equal to
$W=M_{\psi}$ (for comparison with data we identify $2m_c$ with the
mass $M_\psi$ of the $J/\psi$),
\begin{equation}
p_{max}=\frac{1}{2M_B}\left[(M_B^2+m_{sp}^2-M_{\psi}^2)^2
-4m_{sp}^2M_B^2 \right]^\frac{1}{2} \, .
\end{equation}
The first integration in Eq.~(\ref{brac}) results from the transformation from
the spectrum for a $B$ meson at rest to the spectrum for a $B$ meson in flight,
where 
\begin{equation}
k_{\pm}(|{\bf k}_{\psi}|)=\frac{1}{M_B}\left(E_B |{\bf k}_{\psi}|\pm 
|{\bf p}_B| E_{\psi}\right) \, , \hspace{5mm}
\hat{k}_+(|{\bf k}_{\psi}|)=\mbox{min}\{k_+(|{\bf k}_\psi|),\, 
k_{max}\}\, , 
\end{equation}
$k_{max}$ being the maximum value of the $J/\psi$ momentum from the 
decay $B\rightarrow J/\psi+X$ in the restframe of the $B$,
\begin{equation}
k_{max}=\frac{1}{2 M_B}\left[\left(M_B^2+M_\psi^2
-S_{min}\right)^2-4 M_B^2 M_\psi^2\right]^{\frac{1}{2}}\,,
\hspace{1cm} S_{min}=m_{sp}^2\,.
\label{lim}
\end{equation}
(Note that in the PM $S_{min}=0$ for vanishing masses of the final state
quark $q_f$.)

In the following Section we make use of Eq.~(\ref{brac}) to compare the
model predictions with experimental data.
\subsection{Analysis and Numerical Evaluation \label{ac-an}}

Both models, the parton model as well as ACCMM incorporate the bound state
structure of the $B$ meson by postulating a momentum and, consequently, a mass
distribution for the heavy quark. In Section \ref{pman} we pointed out that in 
the PM the dependence of the $J/\psi$ momentum spectrum on the exact value
of the Peterson parameter $\varepsilon_p$ (within the low momentum range 
$|\mbox{\bf k}_\psi|\lesssim 1.4$ GeV) is only moderate. We shall see that,
contrary to this,  in the ACCMM model the dependence of the spectrum on the 
{}Fermi parameter $p_{\SS F}$ is strong, which allows us to perform a detailed
analysis.
 
Introducing in the latter model another $x$-variable as the ratio $x=W/M_B$,
we may calculate the appropriate distribution function $w(x)$ of the $b$ quark 
in the restframe of the $B$ meson as a function of the relative mass $x$.
In Fig.~2b we plot the function $w(x)$ for $m_{sp}=0.15$ GeV and various
values of $p_{\SS F}$, this value for the spectator mass being frequently used 
in studies of the semileptonic decays. Note that the mass distribution shows 
sizable modifications due to the variation of $p_{\SS F}$. Both models have 
the advantage of avoiding the mass of the heavy quark as an independent 
parameter. As a consequence, the phase space is treated correctly because 
they use the mesonic degrees of freedom.

Within the approximations discussed in Section \ref{intro},
the shape of the momentum spectrum in the decay $B\rightarrow J/\psi+X$ 
is determined by the value of $p_{\SS F}$. 
As argued in Ref.~\cite{Kim}, the Fermi momentum parameter
is not a truly free parameter, but it is directly related to the average
kinetic energy of the heavy quark, $\langle \mbox{\bf p}^2\rangle=\frac{3}{2}
p_{\SS F}^2$ through the Gaussian form of the distribution function of 
Eq.~(\ref{ac-dist}). Thus the value of $p_{\SS F}$ can also be deduced 
theoretically from a study of the $b$ quark's average kinetic energy.
Hwang {\it et al.}~\cite{Kim} calculated the latter in the relativistic
quark model, from which they obtained $p_{\SS F}\sim 0.5-0.6$ GeV,
this value being in good agreement with the one deduced from the QCD sum
rule analysis of Ref.~\cite{Ball}, $p_{\SS F}=0.58\pm 0.06$ GeV.
Both results respect the limit $p_{\SS F}\ge 0.49$ GeV which follows from an 
inequality between the expectation value of the $b$ quark's kinetic energy and
that of the chromomagnetic operator derived by Bigi {\it et al.} using the 
methods of the Heavy Quark Effective Theory \cite{Bigi5}. The lower bound from 
the latter inequality, however, could be considerably weakened by higher order 
perturbative corrections \cite{Kap}.

The numerical range of the Fermi motion parameter deduced from the heavy
quark's kinetic energy is in good agreement with a recent study of the
semileptonic $B$ decays: Using dilepton data, the CLEO collaboration has 
performed a model inde\-pen\-dent analysis of the lepton spectrum over 
essentially the full momentum range \cite{vcbcl}. By comparing the results 
with the theoretical prediction of the ACCMM model, Hwang 
{\it et al.}~\cite{Kim} obtained $p_{\SS F}=0.54 {+0.16 \atop -0.15}$ GeV.

Allowing the Fermi momentum to float within the range 
$p_{\SS F}\sim 0.5-0.6$ GeV, we investigate to what extent the $b$ 
quark motion in the $B$ meson can account for the observed spectrum in 
$B\rightarrow J/\psi+X$ decays. One might note that, in contrast to the
energy spectrum in the semileptonic decays, the shape of the $J/\psi$ 
momentum  spectrum is sensitive to the Fermi motion parameter over a wide 
range of the phase space.

Employing the spectator distribution function of Eq.~(\ref{ac-dist}), we 
calculated the momentum spectrum for the decay of $B$ mesons produced at the 
$\Upsilon(4S)$ resonance. Figs.~5 and 6 show the comparison with the CLEO 
data for the set of parameters given in Section \ref{pman} and for 
$m_{sp}=0.15$ GeV. In Fig.~5 one can see
that (considering the fact that the ACCMM model does not include hadronization
effects and, consequently, yields an averaged spectrum) the agreement of the 
theoretical spectrum with the data is good provided we choose the value 
$p_{\SS F}={\cal O}(0.57$ GeV). Again we point out the dominance of the color
octet charmonium production mechanism by plotting also the spectrum which
arises from the color singlet model, the latter underestimating the data by
roughly a factor of three.

To demonstrate the high sensitivity of the spectrum to the Fermi motion
parameter, we present in Fig.~6 the spectrum for $p_{\SS F}=0.3$ GeV (with 
the remaining parameters kept unchanged), the latter value being commonly 
used by experimentalists in the analysis of the semileptonic decays.
Note the large discrepancy between theoretical prediction and data.
Thus it is obvious that the shape of the measured $J/\psi$ momentum 
distribution cannot be reproduced in the model, within the approximations we
made, when using values significantly smaller than $p_{\SS F}=0.5$ GeV. 
Especially the sizable contribution in the low momentum range requires a soft 
probability distribution of the heavy quark momentum. Further we want to 
emphasize that (using $p_{\SS F}=0.57$ GeV) not only
the shape but, due to the inclusion of color octet intermediate states in the 
charmonium production, also the normalization of the theoretical spectrum is 
in good agreement with the data. This way our analysis supports
the numerical values of the NRQCD matrix elements given in Eq.~(\ref{meval}). 

Note that the authors of Ref.~\cite{Ko2} could not account for the
measured value of the branching ratio ${\cal B}(B\rightarrow J/\psi+X)$,
because they restricted the NRQCD analysis to positive values of the matrix
element $\langle O_8^{J/\psi}(^3P_0)\rangle$, and because (adopting 
$m_b=5.3$ GeV) they did not take into account corrections due to the
bound state structure of the $B$ meson.

We conclude that, using the ACCMM model, the Fermi motion of the $b$ quark 
can account for the observed momentum distribution in the 
$B\rightarrow J/\psi+X$ decay accurately. The required value of the Fermi 
momentum $p_{\SS F}={\cal O}(0.57$ GeV) comes out to agree with the range 
deduced from the studies of the heavy quark's kinetic energy, as well as,
with the range recently obtained from the semileptonic $B$ decays. 
\\[12pt]
We close this section with a comment on the distinct 
secondary bump which appears in the low momentum region of the measured 
$J/\psi$ spectrum. While this article was being written, Brodsky and 
Navarra \cite{Br2} suggested an interpretation of the latter in terms of
the three-body decay $B\rightarrow J/\psi\Lambda\bar{p}$, in which they 
assumed that the underlying dynamics of the {\it exclusive} mode is 
reflected in the measured shape of the inclusive momentum distribution. 
The authors argued that the three-body final state accounts for a distinct 
enhancement in the inclusive spectrum, in which the position of the maximum 
($|{\bf k}_\psi|\simeq 0.5$ GeV) coincides with the measured bump. 
Thus our inclusive approach to the $J/\psi$ momentum distribution, when 
combined with the analysis of the exclusive three-body mode 
$J/\psi\Lambda\bar{p}$ \cite{Br2}, yields a spectrum which is in good 
agreement with the measurement over the whole range of phase space.

\section{Summary \label{summary}}
The analysis of the decay $B\rightarrow J/\psi+X$ requires the color singlet
and octet contributions to the intermediate $c\bar{c}$ states. The octet
configurations were found to be necessary for improving the predictions for
the branching ratio \cite{Ko1}. In this article we include the octet
configurations in the analysis of the $J/\psi$ momentum spectrum.
We adopt the leading order expansion of NRQCD, where the kinematics of soft 
gluon radiation associated with the nonperturbative transitions 
$c\bar{c}\rightarrow J/\psi+X$ is neglected.  Thus we address the question 
whether and to what extent the motion of the $b$ quark in the bound state 
can account for the observed momentum spectrum. 
To this end, we review in Eqs.~(\ref{br0})-(\ref{b}) and (\ref{psiten}) the 
changes introduced to the $J/\psi$ tensor through the leading color octet 
configurations. The structure constants of the tensor depend on the NRQCD
matrix elements $\langle O^{J/\psi}_a[n]\rangle$, which parameterize the
conversion of a $c\bar{c}$ pair in angular momentum $[n]$ and color $a$
to the $J/\psi$. Values for the matrix elements are taken from other
experiments. With these results we develop a formalism for the inclusive
decay within the parton and the ACCMM model, respectively. 
The shapes of the spectra originating from various $c\bar{c}(^{2S+1}L_J)_a$ 
configurations are very similar to each other. They are sensitive to the 
motion of the $b$ quark in the $B$ meson. Explicit formulas for the decay 
spectra are given in Eqs.~(\ref{sim}) and (\ref{brac}). They depend on the 
momentum distribution function of the $b$ quark $f(x)$ and the Fermi motion 
$\phi(p)$ of the spectator quark, respectively.  

Numerical calculations in the parton model give a satisfactory presentation
of the data, provided that the heavy quark momentum distribution is taken to 
be soft. To be explicit, we find $\varepsilon_p={\cal O}(0.008-0.012)$ for
the parameter of the Peterson {\it et al.}~distribution function. The ACCMM 
model can account for the observed momentum spectrum more accurately. The
preferred Fermi momentum $p_{\SS F}={\cal O}(0.57$ GeV) is in good agreement 
with the range $p_{\SS F}\sim 0.5-0.6$ GeV deduced from theoretical studies 
of the heavy quark's kinetic energy. 

Our results are obtained in the approximation of neglecting the gluon
momenta at the hadronization stage, i.e., we include only the leading terms 
in the NRQCD expansion. Very recently, it was shown how the resummation of 
higher order nonperturbative corrections to the quarkonium production leads 
to the introduction of universal distribution functions \cite{BRW}. 
It will be a challenge for future studies to investigate the importance of 
these shape functions for the momentum distribution in the decay 
$B\rightarrow J/\psi+X$ and to compare the theory with more accurate data.
%
\vspace*{3cm}
\begin{center}{\large Acknowledgements}
\end{center}
We wish to thank Eric Braaten for helpful comments. 
One of us (PHS) wishes to thank the Deutsche For\-schungs\-gemeinschaft for 
a scholarship in the Graduate Program for Elementary Particle Physics at the 
University of Dortmund.
\newpage
%
%

%
%
\newpage
{\large \bf Figure Captions }
\begin{itemize}
\item[Fig.~1 ]Separation of the various distances contributing to
the $B\rightarrow J/\psi+X$ decays. $Q_{1,\;2}$ are the four-quark operators
which induce the effective $b\rightarrow c\bar{c}q_f$ transitions.
\item[Fig.~2a]Momentum distribution function $f(x)$ in the PM for various 
values of $\varepsilon_p$.
\item[Fig.~2b]Mass distribution function $w(x)$ in the ACCMM model for
various values of $p_{\SS F}$.
\item[Fig.~3 ]Theoretical momentum spectrum in the PM for direct inclusive
$J/\psi$ production from $B$ decays at the $\Upsilon(4S)$ resonance, 
shown for various values of $\varepsilon_p$ and compared with the CLEO 
data; the parameters are given in Section \ref{pman}.
\item[Fig.~4 ]Same as in Fig.~3, now for $\varepsilon_p=0.012$ and a modified 
value of the $(^3P_1)$ matrix element (dashed line). The dotted line shows 
the contribution from color singlet $(^3S_1)$ intermediate states.  
\item[Fig.~5 ]Theoretical momentum spectrum in the ACCMM model for 
inclusive $B$ decays to $J/\psi$ with $p_{\SS F}=0.57$ GeV, 
compared with the CLEO data; the parameters are given in Section \ref{pman}.
The dashed line shows the contribution from color singlet $(^3S_1)$
intermediate states.
\item[Fig.~6 ]Same as in Fig.~5, now the theoretical momentum spectrum 
in the ACCMM model for various values of $p_{\SS F}$ as shown.
\end{itemize}
%
\newpage
\vspace*{6.4cm}
\bf
\large
\noindent
\hspace*{1cm}\epsfig{file=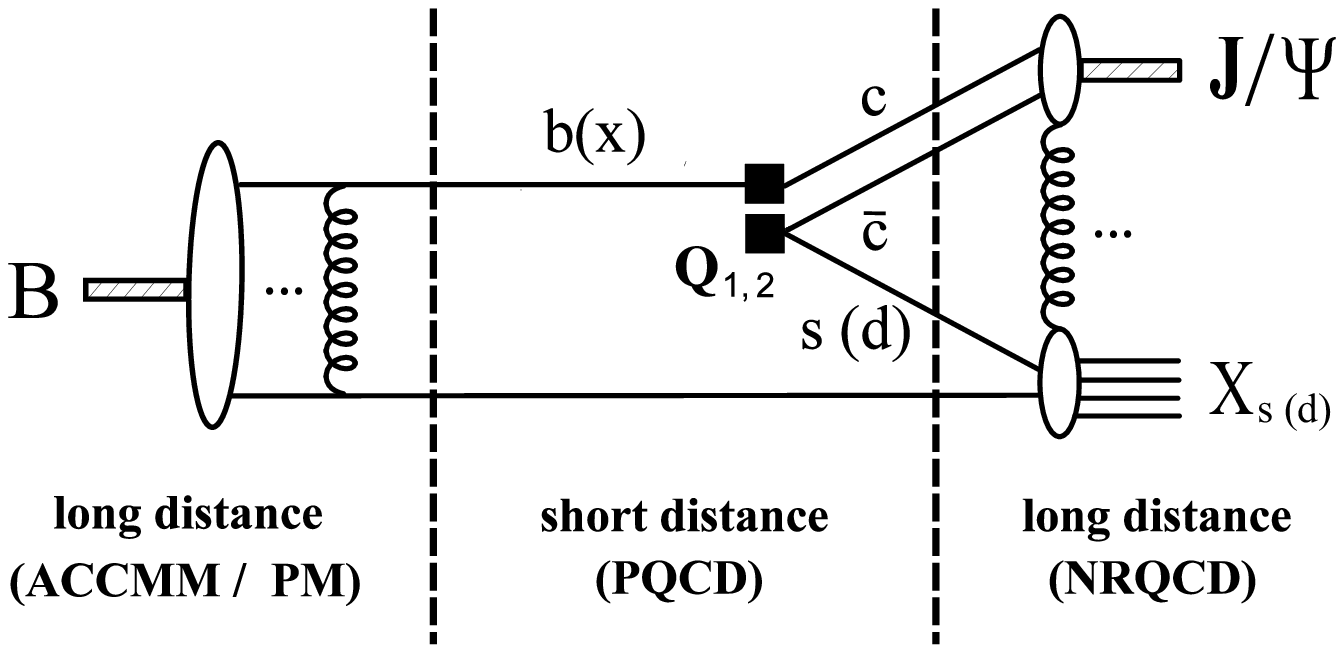,width=13.5cm}
\vspace*{1cm}\\
\centerline{fig. 1}
\newpage
\noindent
\hspace*{1.18cm}\epsfig{file=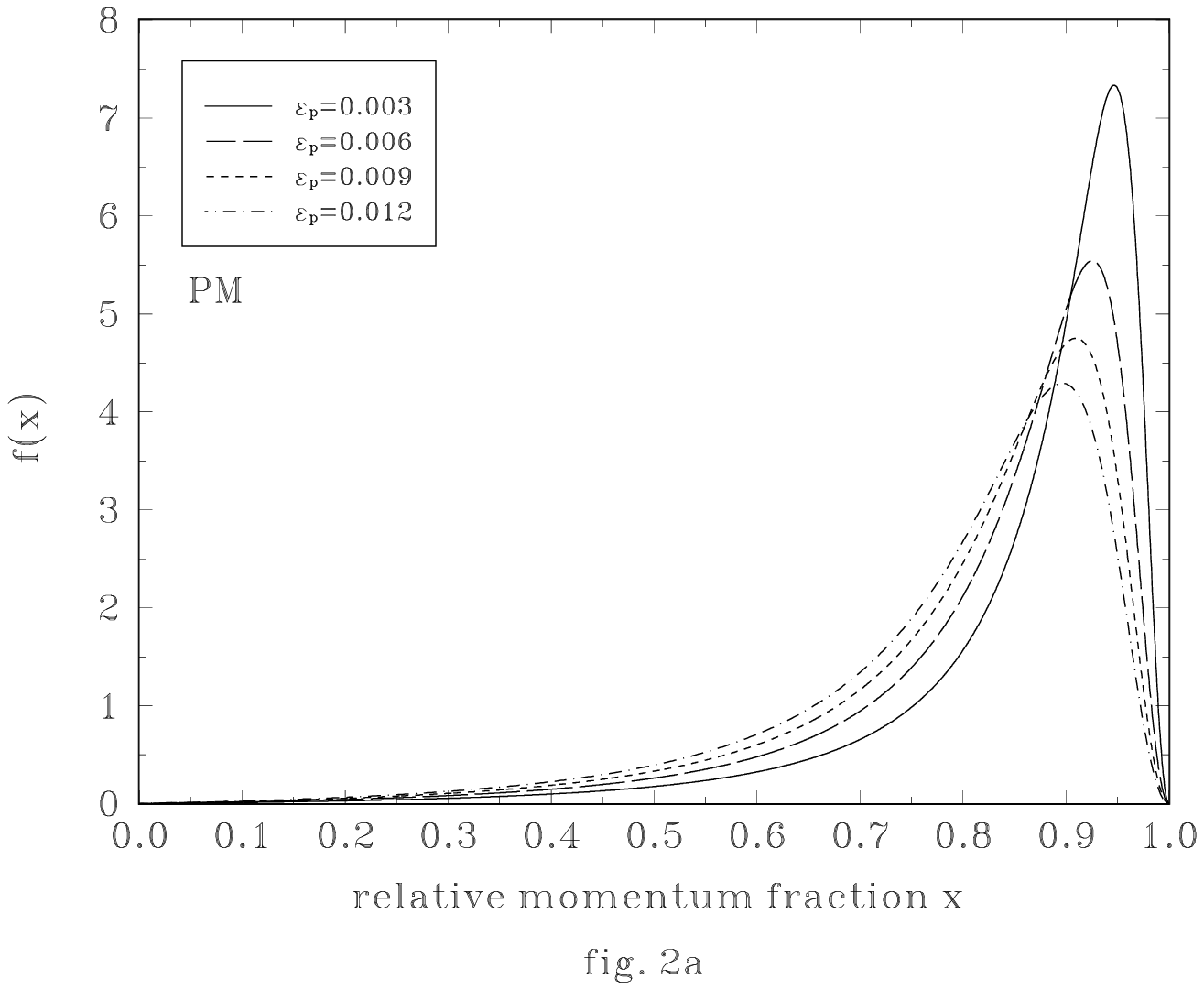,height=11cm,angle=360}\\
\vspace*{-1cm}
\noindent
\hspace*{1cm}\epsfig{file=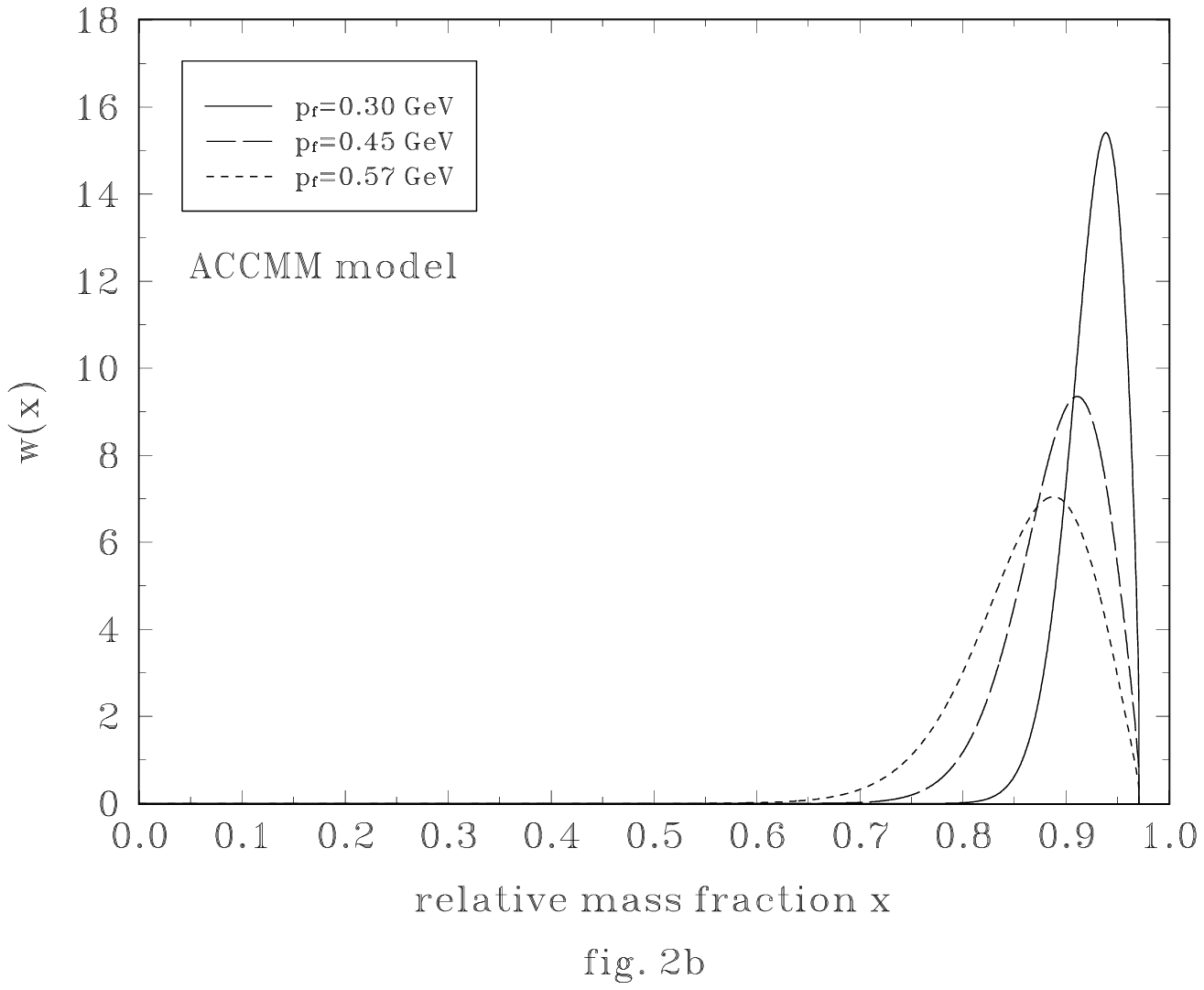,height=11cm,angle=360}
\newpage
\noindent
\hspace*{1.18cm}\epsfig{file=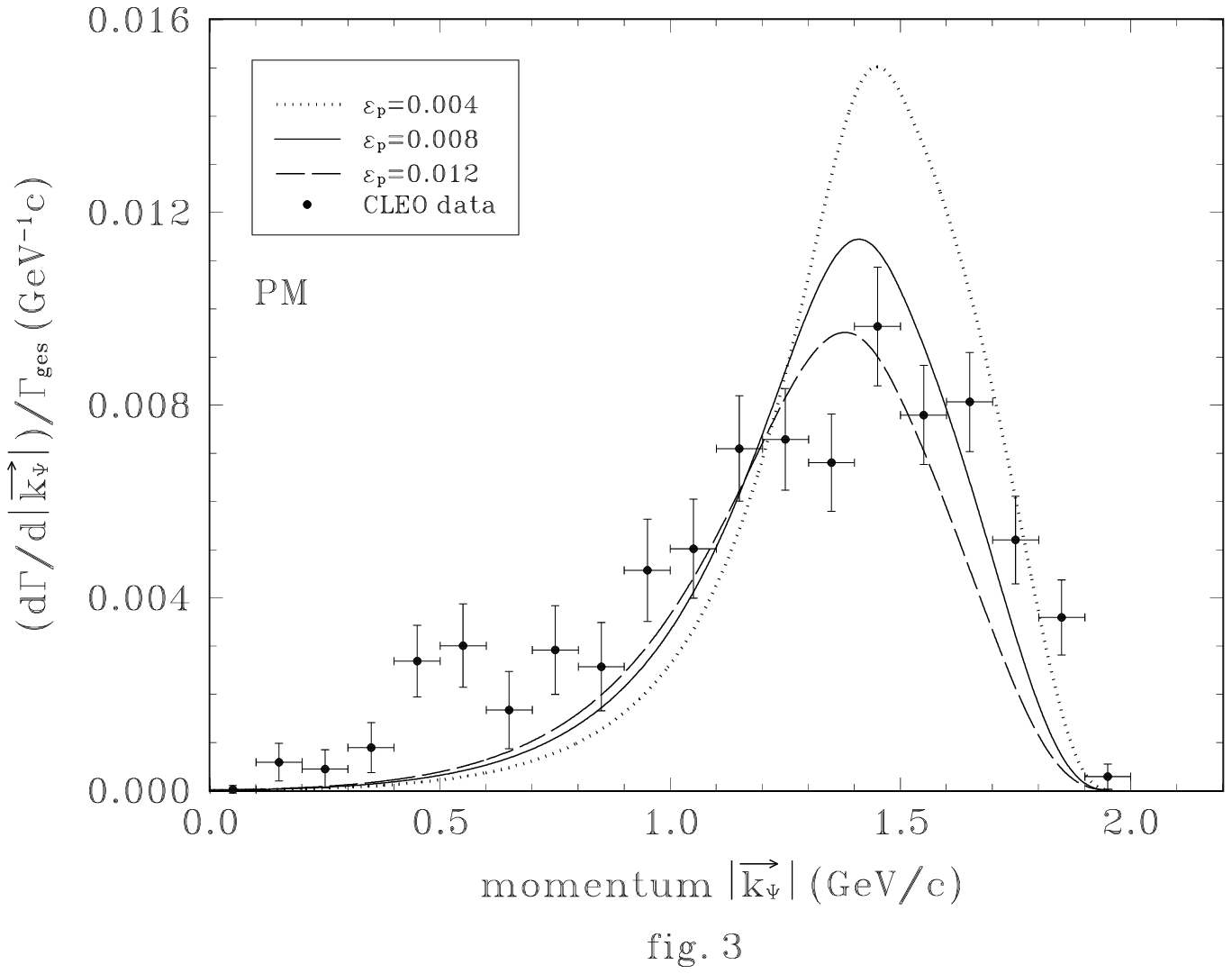,height=11cm,angle=360}\\
\vspace*{-1cm}
\noindent
\hspace*{1cm}\epsfig{file=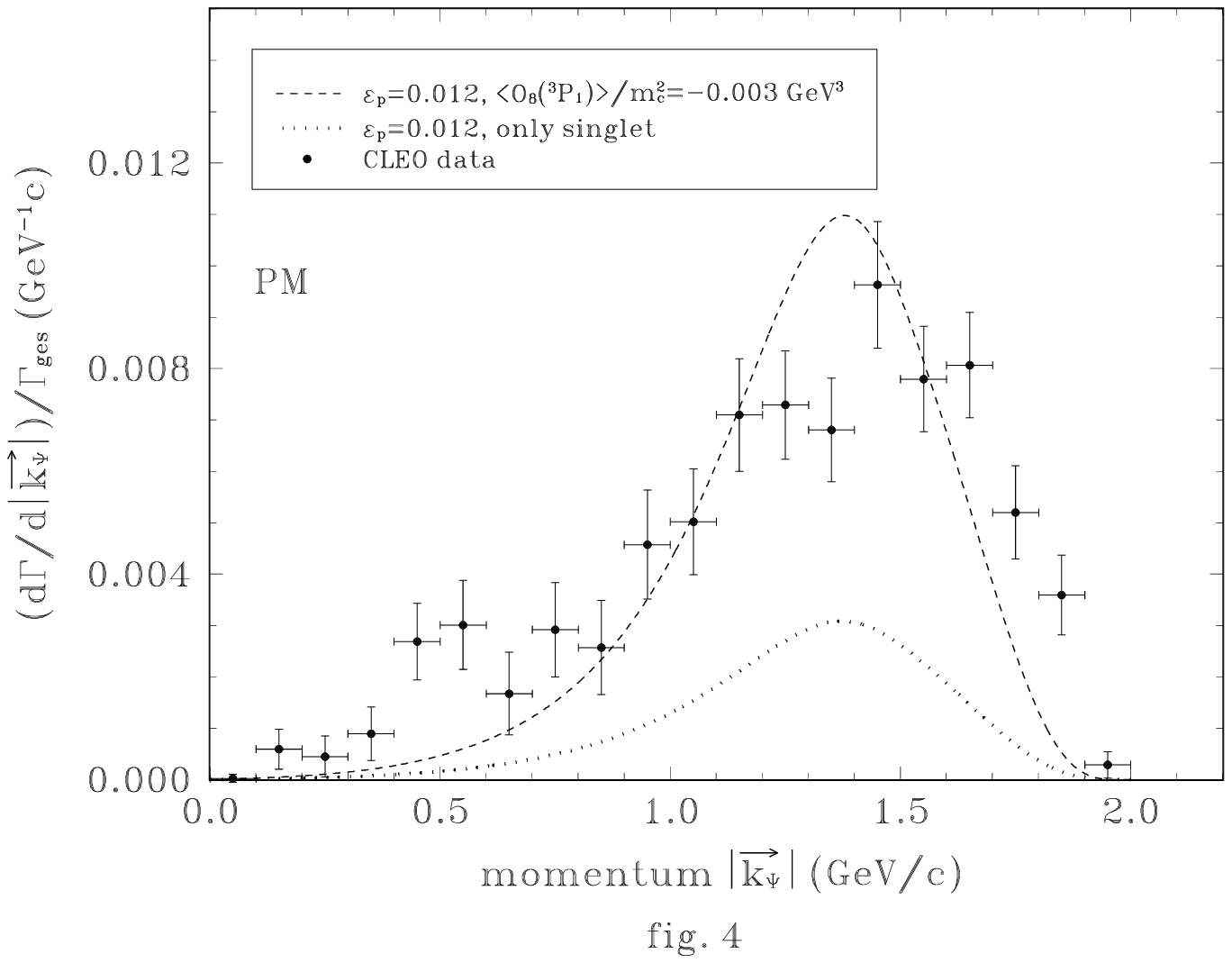,height=11cm,angle=360}
\newpage
\noindent
\hspace*{1.18cm}\epsfig{file=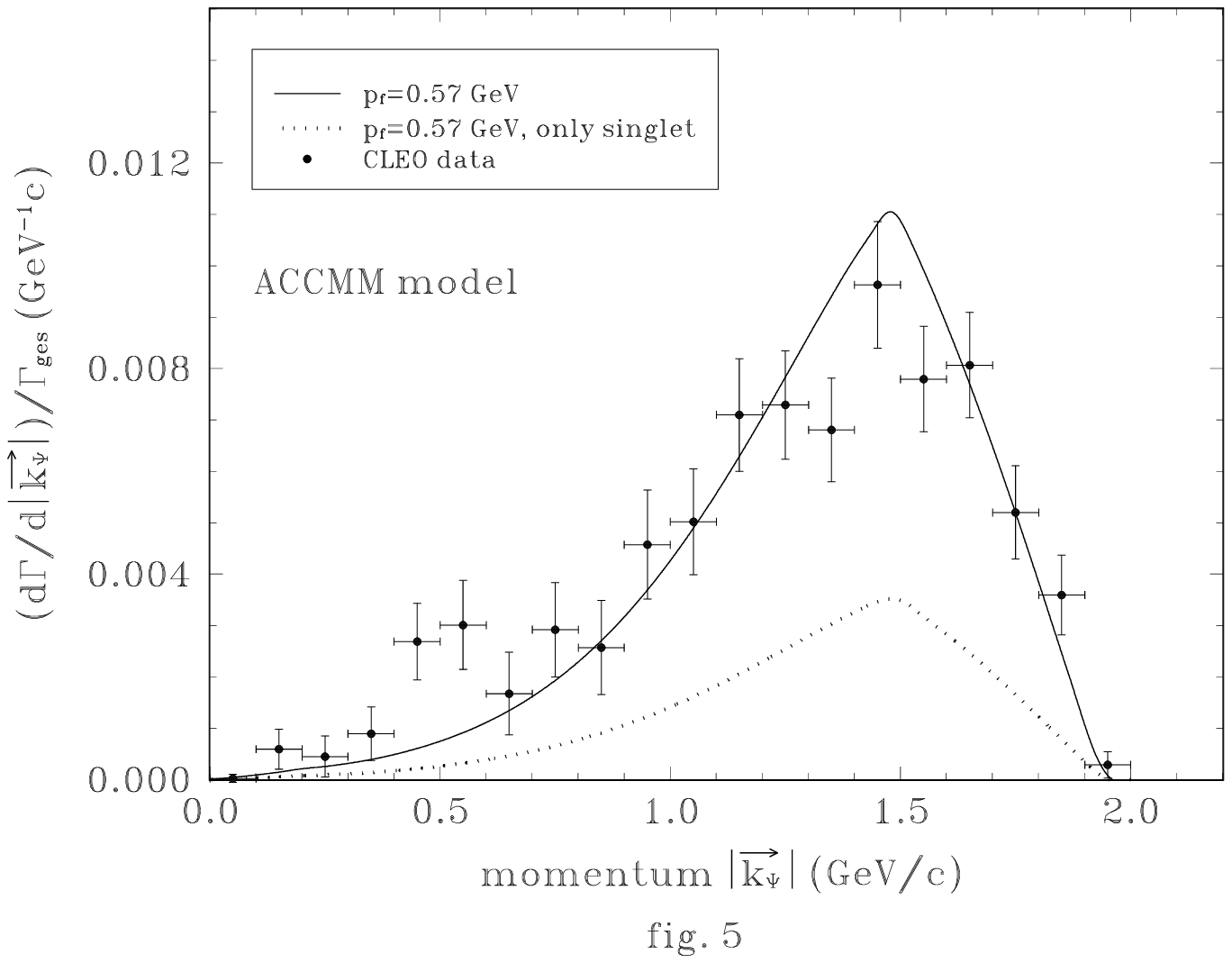,height=11cm,angle=360}\\
\vspace*{-1cm}
\noindent
\hspace*{1cm}\epsfig{file=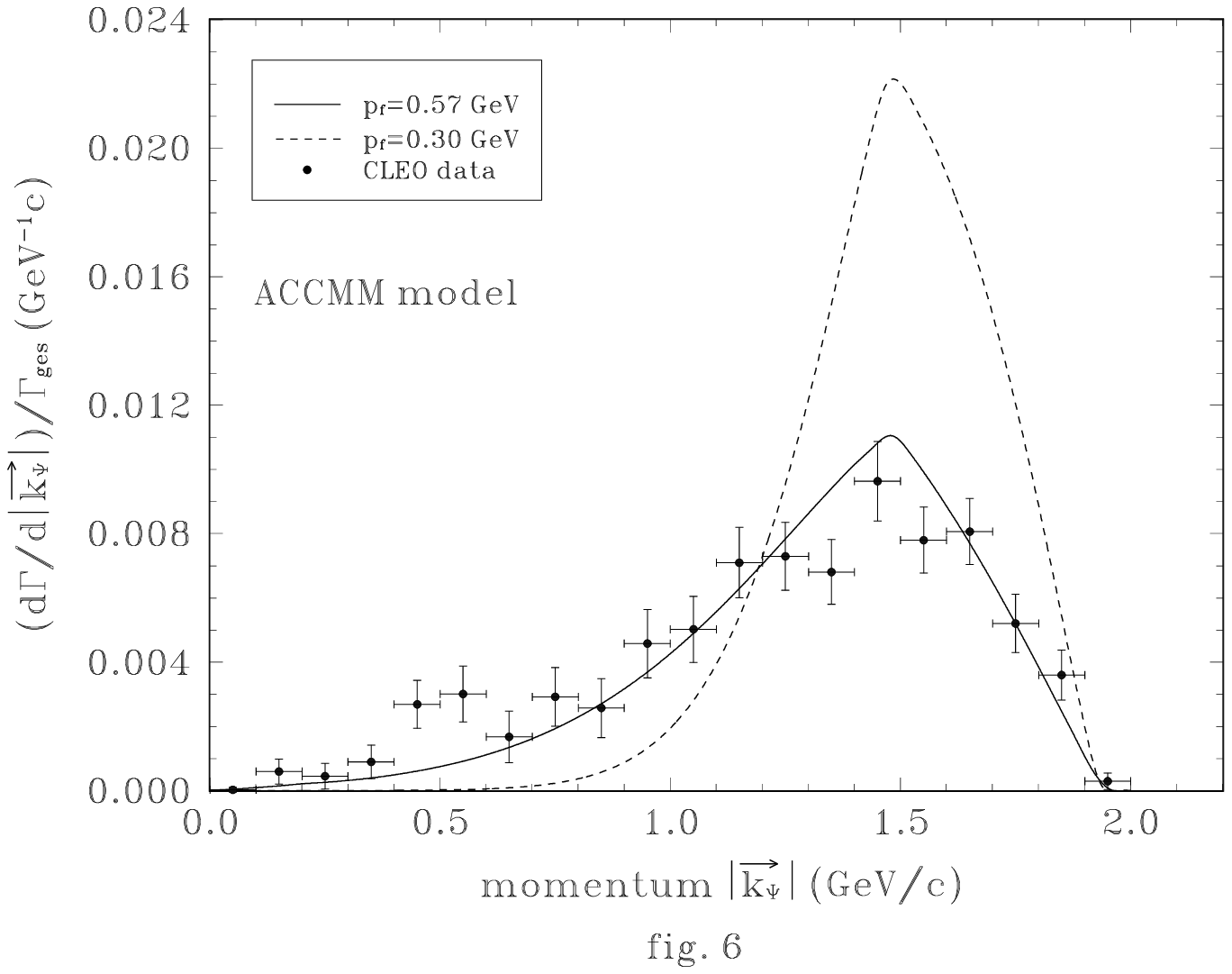,height=11cm,angle=360}

\begin{thebibliography}{99}
%
\bibitem{AaC}
H. Albrecht {\it et al.} (ARGUS Collaboration), Phys. Lett. {\bf B 199}, 
(1987) 451; D. Bortoletto {\it et al.} (CLEO Collaboration), 
Phys. Rev. {\bf D 45}, (1992) 21.
\bibitem{Cleo1}
R. Fulton {\it et al.} (CLEO Collaboration), Phys. Rev. {\bf D 52},
(1995) 2661.
\bibitem{csin}
M.B. Wise, Phys. Lett. {\bf B 89}, (1980) 229;
J.H. K\"uhn, S. Nussikov, and R. R\"uckl, Z. Phys. {\bf C 5}, (1980) 117;
J.H. K\"uhn and R. R\"uckl, Phys. Lett. {\bf B 135} (1984) 477;
Erratum-ibid. {\bf 258}, (1991) 499;
G. Bodwin, E. Braaten, T.C. Yuan, and G.P. Lepage, Phys. Rev. {\bf D 46},
(1992) 3703.
\bibitem{Berg}
L. Bergstr\"om and P. Ernstr\"om, Phys. Lett. {\bf B 328}, (1994) 153.
\bibitem{SoTo}
J.M. Soares and T. Torma, preprint UMHEP-438 (Feb 97), to be published
in Phys. Rev. D.
\bibitem{BBL}
G.T. Bodwin, E. Braaten, and G.P. Lepage, Phys. Rev. {\bf D 51}, (1995)
1125.
\bibitem{BrFl}
E. Braaten and S. Fleming, Phys. Rev. Lett. {\bf 74}, (1995) 3327.
\bibitem{ChL}
P. Cho and A.~K. Leibovich, Phys. Rev. {\bf D 53}, (1996) 150;
Phys. Rev. {\bf D 53}, (1996) 6203.
\bibitem{FM}
S. Fleming and I. Maksymyk, Phys. Rev. {\bf D 54} (1996) 3608.
\bibitem{BRGS}
M. Beneke and I.Z. Rothstein, Phys. Rev. {\bf D 54}, (1996) 2005;
Erratum-ibid. {\bf D 54}, 7082;
S. Gupta and K. Sridhar, Phys. Rev. {\bf D 54}, (1996) 5545;
Phys. Rev. {\bf D 55}, (1997) 2650.
\bibitem{BrCh}
E. Braaten and Y.-Q. Chen, Phys. Rev. Lett. {\bf 76}, (1996) 730.
\bibitem{CaKr}
M. Cacciari and M. Kr\"amer, Phys. Rev. Lett. {\bf 76}, (1996) 4128.
\bibitem{Am}
J. Amundson, S. Fleming, and I. Maksymyk, preprint UTTG-10-95,
MADTH-95-914 (Jan 1996).
\bibitem{Ko2}
P. Ko, J. Lee, and H.S. Song, Phys. Rev. {\bf D 54}, (1996) 4312.
\bibitem{Z0}
K. Cheung, W.-Y. Keung, and T.C. Yuang, Phys. Rev. Lett. {\bf 76}, (1996)
877; P. Cho, Phys. Lett {\bf B 368}, (1996) 171; S. Baek, P. Ko, J. Lee,
and H.S. Song, Phys. Lett. {\bf B 389}, (1996) 609.
\bibitem{Fl}
S. Fleming, talk at the Quarkonium Physics Workshop: Experiment Confronts
Theory, Chicago, IL, June 1996 (MADPH-96-966).
\bibitem{ccrev}
E. Braaten, S. Fleming, and T.C. Yuan,
Ann. Rev. Nucl. Part. Sci. {\bf 46}, (1996) 197.
\bibitem{Ko1}
P. Ko, J. Lee, and H.S. Song, Phys. Rev. {\bf D 53}, (1996) 1409.
\bibitem{FHMN}
S. Fleming, O.F. Hern{\'a}ndez, I. Maksymyk, and H. Nadeau,
Phys. Rev. {\bf D 55}, (1997) 4098.
\bibitem{Pal-St}
W.F. Palmer and B. Stech, Phys. Rev. {\bf D 48}, (1993) 4174.
\bibitem{hqope}
J. Chay, H. Georgi, and B. Grinstein, Phys. Lett. {\bf B 247}, (1990) 399;
I. Bigi and N.G. Uraltsev, Phys. Lett. {\bf B 280}, (1992) 120;
I. Bigi, N.G. Uraltsev, and A. Vainshtein, Phys. Lett. {\bf B 293}, (1992)
430; Erratum-ibid. {\bf 297}, (1993) 477;
I. Bigi, M. Shifman, N.G. Uraltsev, and A. Vainshtein, Phys. Rev. Lett. 
{\bf 71}, (1993) 496;
M. Neubert, Phys. Rev. {\bf D 49}, (1994) 1542;
T. Mannel, Proceedings of the 138th WE-Heraeus Seminar, edited by J.G. 
K\"orner and P. Kroll (World Scientific, 1995).
\bibitem{HQET}
{}For reviews see:\\
T. Mannel, in  {\it QCD- 20 Years Later}, edited by P.M. Zerwas and H.A.
Kastrup (World Scientific, Singapore, 1993);
M. Neubert, Phys. Rept. {\bf 245}, (1994) 259.
\bibitem{BRW}
M. Beneke, I.Z. Rothstein, and M.B. Wise, preprint hep-ph/9705286.
\bibitem{Alta}
G. Altarelli, N. Cabibbo, G. Corb\`{o}, L. Maiani, and G. Martinelli,
Nucl. Phys. {\bf B 208}, (1982) 365.
\bibitem{Buras}
A. Buras and P.H. Weisz, Nucl. Phys. {\bf B 333}, (1990) 66.
\bibitem{HS}
X.-G. He and A. Soni, Phys. Lett. {\bf B 391}, (1997) 456.
\bibitem{Jin1}
C.H. Jin, W.F. Palmer, and E.A. Paschos, preprint DO-TH 93/21 and
OHSTPY-HEP-T-93-011, (1993) (unpublished);
C.H. Jin, W.F. Palmer, and E.A. Paschos, Phys. Lett. {\bf B 329}, (1994) 364.
\bibitem{Jin3}
C.H. Jin and E.A. Paschos, preprint DO-TH 95/07. 
\bibitem{mb}
R.M. Barnett {\it et al.} (Particle Data Group), Phys. Rev. {\bf D 54}, 
(1996) 1.
\bibitem{tB}
S. Komamiya, invited talk at the International Europhysics Conference on
High Energy Physics, Brussels, Belgium, July 27 - August 2 1995,
published in the proceedings ({\it Brussels EPS HEP 1995}, 727).
\bibitem{vcbcl}
B. Barish {\it et al.} (CLEO Collaboration), Phys. Rev. Lett. {\bf 76},
(1996) 1570.
\bibitem{ISGW}
N. Isgur, D. Scora, B. Grinstein and M.B. Wise, Phys. Rev. {\bf D 39},
(1989) 799.
\bibitem{vcbav}
S. Stone, talk at the NATO Advanced Study Institute on Techniques and
Concepts of High Energy Physics, Virgin Islands, July 1996 (HEPSY-96-01).
\bibitem{BSK}
G.T. Bodwin, D.K. Sinclair, and S. Kim, Phys. Rev. Lett. {\bf 77}, 
(1996) 2376.
\bibitem{BjBr}
J.D. Bjorken, Phys. Rev. {\bf D 17}, (1978) 171;
S.J. Brodsky, C. Peterson, and N. Sakai, Phys. Rev. {\bf D 23}, (1981) 2745.
\bibitem{PCB}
C. Peterson, D. Schlatter, J. Schmitt, and P.M. Zerwas, Phys. Rev. {\bf D 27},
(1983) 105; J. Chrin, Z. Phys. {\bf C 36}, (1987) 163;
D. Bortoletto {\it et al.} (CLEO Collaboration),
Phys. Rev. {\bf D 37}, (1988) 1719; Erratum-ibid. {\bf 39} (1989), 1471.
\bibitem{Barger}
V. Barger, W.Y. Keung, J.P. Leveille, and R.J.N. Phillips, Phys. Rev. 
{\bf D 24}, (1981) 2016.
\bibitem{Kim}
D.S. Hwang, C.S. Kim, W. Namgung, Z. Phys. {\bf C 69}, (1996) 107;
Phys. Rev. {\bf D 54}, (1996) 5620.
\bibitem{Ball}
E. Bagan, P. Ball, and P. Gosdzinsky, Phys. Lett. {\bf B 342}, (1995) 362.
\bibitem{Bigi5}
I. Bigi, M. Shifman, N.G. Uraltsev, and A. Vainshtein,
Phys. Lett. {\bf B 328}, (1994) 431; Int. J. Mod. Phys. {\bf A 9}, (1994) 
2467, Phys. Rev. {\bf D 52}, (1995) 196.
\bibitem{Kap}
A. Kapustin, Z. Ligeti, M.B. Wise, and B. Grinstein, Phys. Lett.
{\bf B 375}, (1996) 327.
\bibitem{Br2}
S. J. Brodsky and F.S. Navarra, preprint SLAC-PUB-7445 (Apr 1997).
%
\end{thebibliography}
\end{document}